\documentclass[fleqn,10pt]{SelfArx} 

\definecolor{color1}{RGB}{0,0,90} 
\definecolor{color2}{RGB}{0,20,20} 

\usepackage{hyperref} 
\hypersetup{hidelinks,colorlinks,breaklinks=true,urlcolor=color2,citecolor=color1,linkcolor=color1,bookmarksopen=false,pdftitle={Title},pdfauthor={Author},urlcolor=blue}

\usepackage{graphicx}
\usepackage[square]{natbib}
\usepackage{amsmath}
\usepackage{multirow}
\usepackage{subfigure}

\JournalInfo{Published in Computational Materials Science, Vol. 82 (1), 293-297, 2014} 
\Archive{\href{http://dx.doi.org/10.1016/j.commatsci.2013.09.051}{DOI: 10.1016/j.commatsci.2013.09.051}} 

\PaperTitle{Strain in the mesoscale kinetic Monte Carlo model for sintering} 

\Authors{R. Bj\o{}rk$^1$, H. L. Frandsen$^1$, V. Tikare$^2$, E. Olevsky$^{3}$, N. Pryds$^1$} 
\affiliation{\textsuperscript{1}\textit{Department of Energy Conversion and Storage, Technical University of Denmark - DTU, Frederiksborgvej 399, DK-4000 Roskilde, Denmark}} 
\affiliation{\textsuperscript{2}\textit{Sandia National Laboratory, Albuquerque, NM 87185, New Mexico, USA}} 
\affiliation{\textsuperscript{3}\textit{Mechanical Engineering Department, San Diego State University, 5500 Campanile Dr., San Diego, CA 92182-1323, USA}} 
\affiliation{*\textbf{Corresponding author}: rabj@dtu.dk} 

\Keywords{} 
\newcommand{\keywordname}{Keywords} 

\Abstract{Shrinkage strains measured from microstructural simulations using the mesoscale kinetic Monte Carlo (kMC) model for solid state sintering are discussed. This model represents the microstructure using digitized discrete sites that are either grain or pore sites. The algorithm used to simulate densification by vacancy annihilation removes an isolated pore site at a grain boundary and collapses a column of sites extending from the vacancy to the surface of sintering compact, through the center of mass of the nearest grain. Using this algorithm, the existing published kMC models are shown to produce anisotropic strains for homogeneous powder compacts with aspect ratios different from unity. It is shown that the line direction biases shrinkage strains in proportion the compact dimension aspect ratios. A new algorithm that corrects this bias in strains is proposed; the direction for collapsing the column is determined by choosing a random sample face and subsequently a random point on that face as the end point for an annihilation path with equal probabilities. This algorithm is mathematically and experimentally shown to result in isotropic strains for all samples regardless of their dimensions. Finally, the microstructural evolution is shown to be similar for the new and old annihilation algorithms.}

\begin{document}

\flushbottom 

\maketitle 


\thispagestyle{empty} 

\section{Introduction}
Sintering is an important high temperature densification process that relies on diffusion processes to form e.g. dense microstructures with  high mechanical strength. Almost all ceramics and many metals used in a number of technologies are fabricated from compacted powder particles that are sintered. Therefore, it is important to understand and determine the final sintered microstructure and shrinkage strains as a function of the green powder compact variables such as specific grain size and pore size distribution.

\begin{figure*}[!t]
  \centering
\includegraphics[width=1.5\columnwidth]{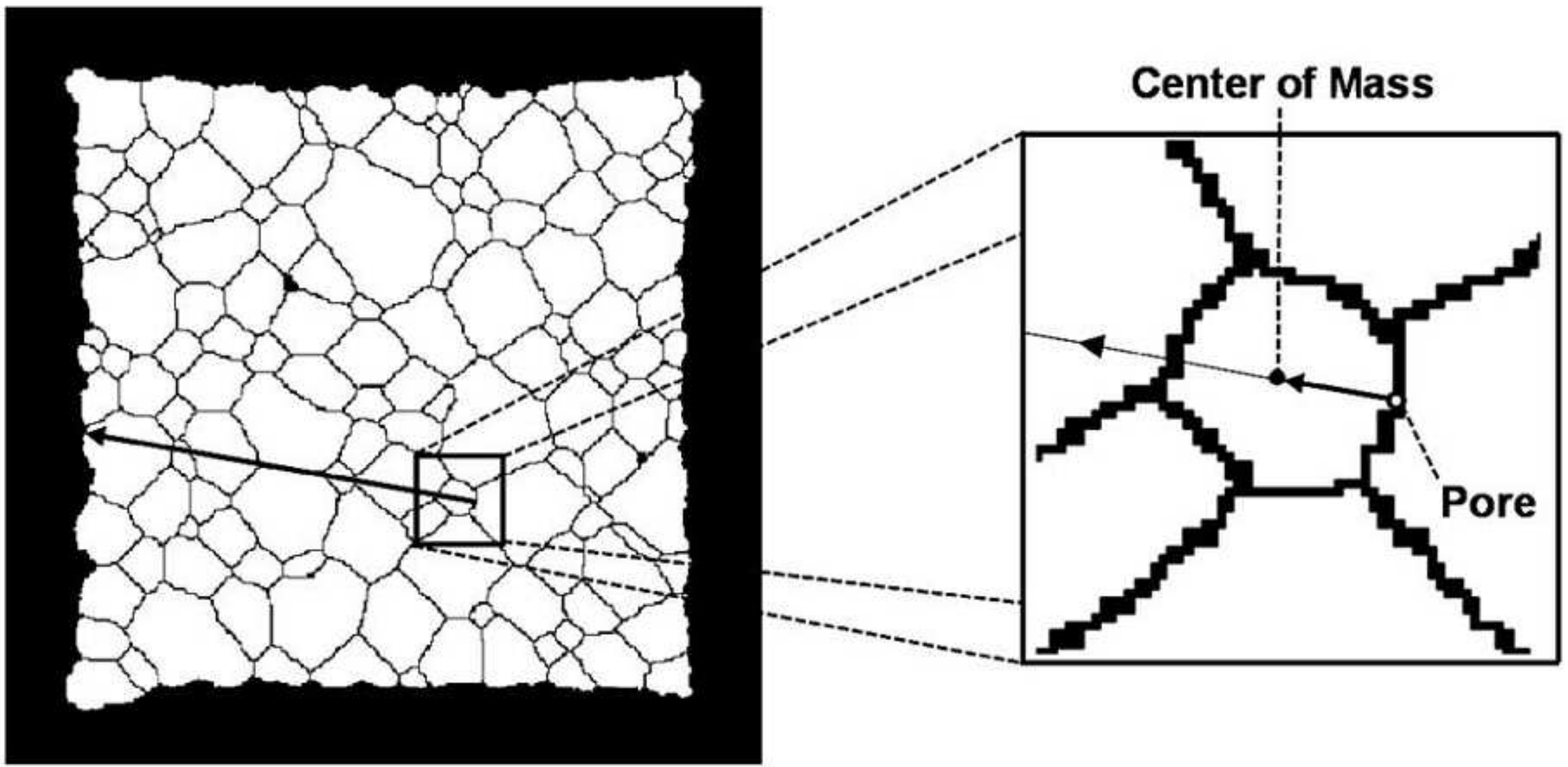}
\caption{Schematic of the annihilation algorithm with collapsing column. Black color denotes pores and grain boundaries; white denotes grains. The direction of annihilation is indicated. From Braginsky et. al. 2005 \cite{Braginsky_2005}. Copyright Elsevier.}
    \label{Fig_Braginsky_2005_Fig_1}
\end{figure*}

Traditionally, microstructural investigation of sintered samples has been conducted experimentally using scanning electron microscopy on fractured or cross sectioned sintered specimens. However, in recent years simulations of the microstructural evolution during sintering have received increasing scientific interest. Such a modelling approach can be advantageous as pure samples with no defects or agglomerations can be studied. Simulating microstructural evolution during sintering has been conducted using three different approaches. Wakai et. al. \cite{Wakai_2003, Wakai_2003b, Wakai_2011,Wakai_2012} have used Surface Evolver to track the motion of pore surfaces and grain boundaries in response to local curvatures. This technique allows the microstructural evolution to be followed with extreme detail, but only so, for a single digit number of grains and with limited insight on the grain growth and other mesoscale phenomena. Parhami \& McMeeking \cite{Parhami_1994}, Wonish et. al. \cite{Wonisch_2009}, Martin \& Bordia \cite{Martin_2009}, Olmos et. al. \cite{Olmos_2009} and Rasp et. al. \cite{Rasp_2012} among other have used a discrete element method (DEM) approach, in which individual particles are modeled as spheres, whose motion are governed by Newtonian mechanics. The spherical particles are allowed to overlap to simulate densification. This allows large samples to be modeled during the early stages of sintering, but grain coarsening is usually not included in these models.  Finally, Tikare et. al. \cite{Tikare_2003,Braginsky_2005,Tikare_2010,Cardona_2011,Cardona_2012, Bjoerk_2012a, Bjoerk_2012b}, Matsubara et. al. \cite{Matsubara_1999,Kishino_2002,Mori_2004,Matsubara_2005} and Zhang et. al. \cite{Zhang_2003} have used a kinetic Monte Carlo (kMC) model to simulate solid-state sintering of both single particles and large powder compacts. In this model the microstructure is digitized and evolved on a cubic grid based on the local curvature of pores. All materials processes such as densification, grain growth, and grain and pore shape changes are modeled by direct simulation of diffusion processes based on temperature. A finite element form of the Monte Carlo approach has also been used by Bord{\`e}re et. al. \cite{Bordere_2002,Bordere_2006,Bordere_2008} to model viscous sintering.

So far the microstructural properties of the studied samples, such as pore and grain size, have been considered extensively in the different models. However, the macroscopic properties of the simulated powder compacts, such as the shrinkage strain of the sample, have been largely ignored. While the DEM approach allows for direct calculation of strains on a sample, the shrinkage strain has until now rarely been considered in simulations using the kinetic Monte Carlo approach. This is partly because mostly quadratic or cubic samples have been considered \cite{Tikare_2005}. Experimentally, freely sintering homogeneous samples show isotropic strain regardless of sample dimensions \cite{Ozer_2006,Guillon_2007a,Liu_2007}, and this should of course be replicated by simulations. Here we investigate the strain of a freely sintering sample modeled using the kMC model for sintering. We consider the specific implementation of the kMC model presented in Tikare et. al. (2010) \cite{Tikare_2010}, but the described approach applies to all kMC sintering models with annihilations.

\begin{figure}[!b]
  \centering
\includegraphics[width=1\columnwidth]{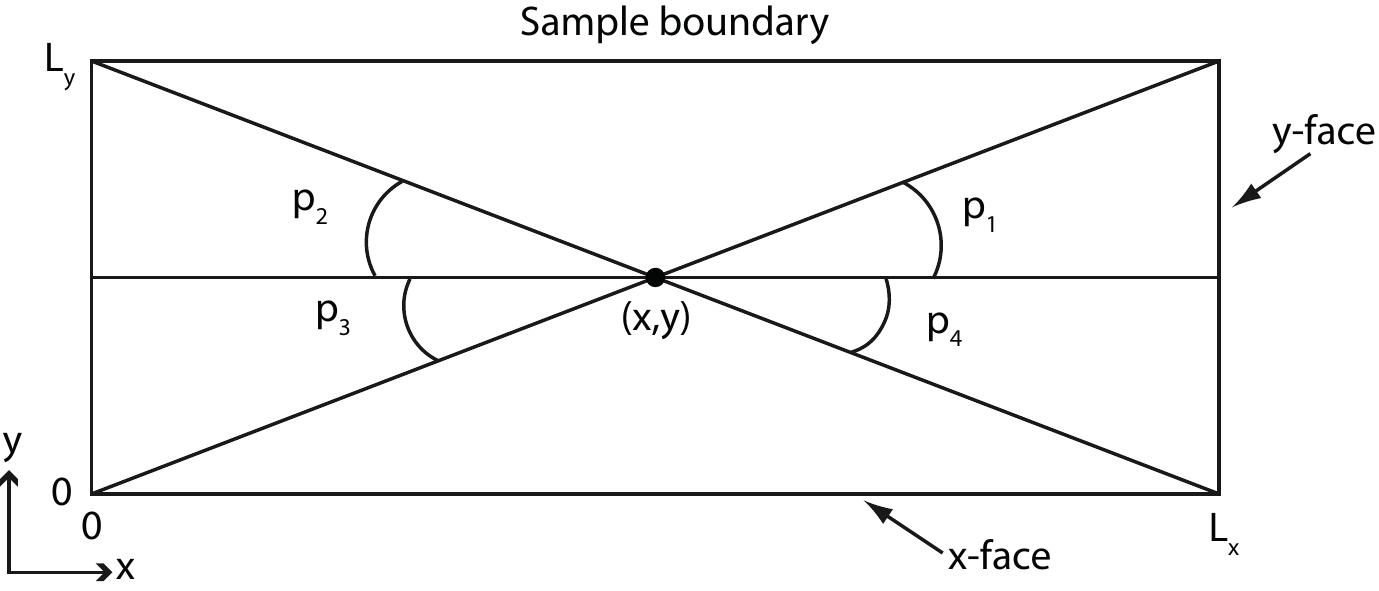}
\caption{The geometry and angles for a box with dimensions $L_x$ and $L_y$. The ``radiation'' angles, $p_{1-4}$ are also illustrated for the point $(x,y)$.}
    \label{Fig_Angles}
\end{figure}

\section{Strain in the current kMC model}
The kMC sintering model is described in detail in the works mentioned above, and only the densification algorithm will be reviewed here as it is pertinent to the shrinkage. The microstructure in a kMC model consists of pores and grains digitized on a cubic grid. Densification takes place when an isolated pore site, called a vacancy, is annihilated by moving the vacancy to the surface of the sample by collapsing a column of single sites extending from vacancy to the surface of the powder compact. Once a vacancy has been selected for annihilation, the column of sites is selected by drawing a straight line from the vacancy through the center of mass (CM) of the adjacent grain to the surface of the sample. This collapsing algorithm is illustrated in Fig. \ref{Fig_Braginsky_2005_Fig_1}. For a cubic sample this will produce isotropic strains, but so far the strain have not been considered for samples with aspect ratios different from one. Here we consider rectangular samples in two dimensions and cuboid\footnote{Also known as rectangular parallelepiped or simply rectangular box.} samples in three dimensions.

Since the direction from a chosen pore site to the center of mass of the adjacent grain will be uniformly distributed with equal frequency in all direction when a large number of grains and annihilations are considered, the present model of annihilation can be compared with a model in which the direction of annihilation is chosen at random with equal probability in all directions. For a two dimensional rectangular sample, the random annihilation direction model can be constructed by picking a random site in the sample, selecting a random direction and finding the surface in that direction. The shrinkage strain is that direction is than increased by an amount of 1/A, where A is the length of that edge in two dimensions and area of that surface in three dimensions. Consider a given random site ($x$,$y$) in a rectangular sample that extends from $x=0$ to $x=L_x$ and $y = 0$ to $y = L_y$, and consists of a porous microstructure with of equiaxed grains with uniformaly distributed intergranular porosity. Vacancies can occur on any grain boundary anywhere in the sample. The probability that an annihilation path will terminate at either of the two $y$-faces can be computed from the ``radiation'' angles, illustrated in Fig. \ref{Fig_Angles},
\begin{eqnarray}\label{Eq.Def_prop}
p_1 & = & \mathrm{atan}\frac{L_y-y}{L_x-x}\\\nonumber
p_2 & = & \mathrm{atan}\frac{y}{L_x-x}\\\nonumber
p_3 & = & \mathrm{atan}\frac{L_y-y}{x}\\\nonumber
p_4 & = & \mathrm{atan}\frac{y}{x}
\end{eqnarray}
with the total probability that the annihilation path passes through the $y$-faces given by $P = (p_1+p_2+p_3+p_4)/(2\pi)$, as the circumference of the full circle is $2\pi$. This local probability can be calculated for every point in the sample. By selecting a large number of annihilation events at random locations and directions, the number of termination points for annihilation on $x$- or $y$-faces can be computed and from this the strain can be calculated.

\begin{figure}[!t]
  \centering
\includegraphics[width=1\columnwidth]{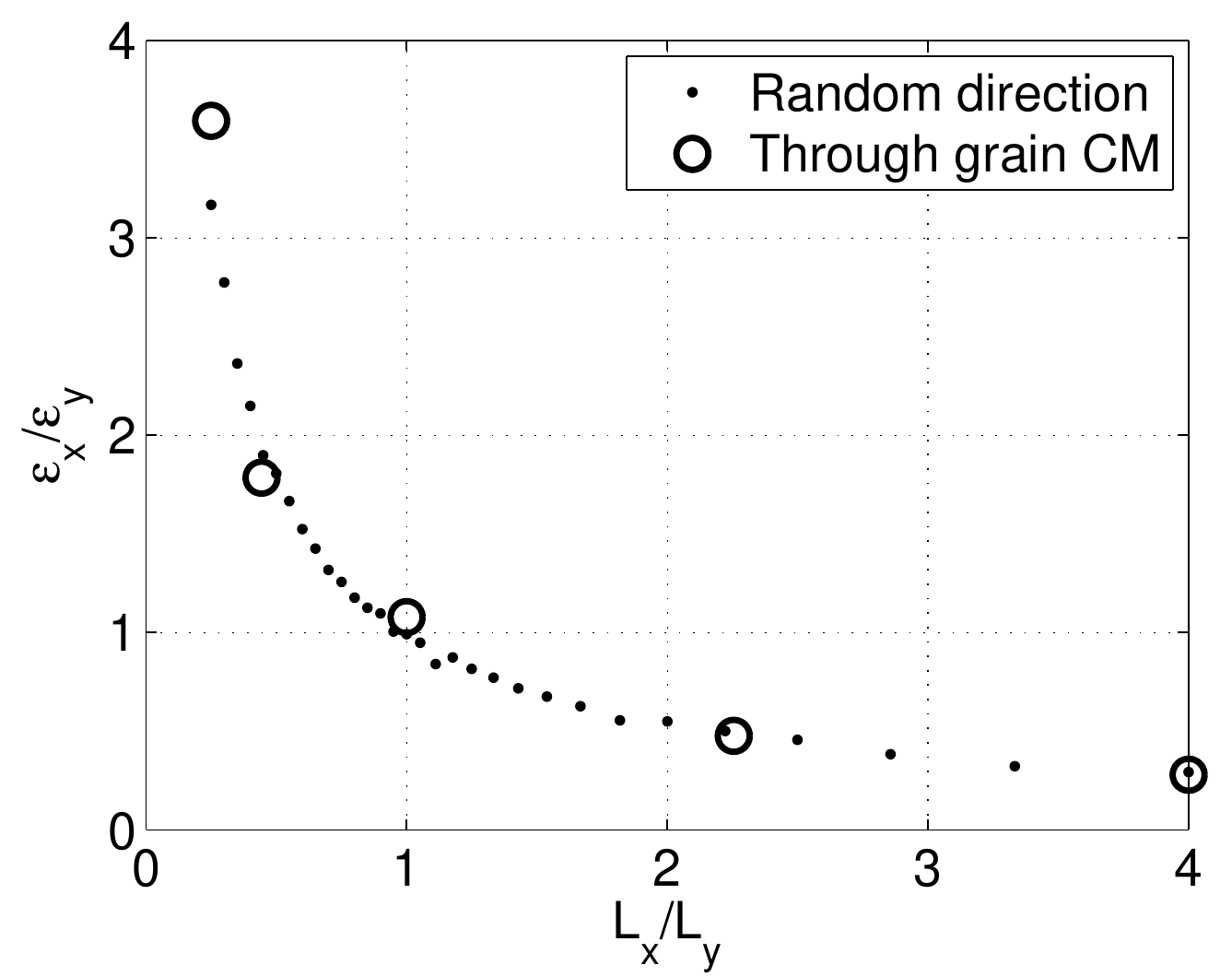}
\caption{The ratio between the shrinkage strain, $\epsilon$, in the $x$ and $y$ directions as function of the aspect ratio of the sample, i.e. $L_x$/$L_y$.}
    \label{Fig_Strain_random_dir_vs_grain_CM}
\end{figure}

Shown in Fig. \ref{Fig_Strain_random_dir_vs_grain_CM} is the ratio between the strain in the $x-$ and $y-$ direction computed from this random model, based on 5000 points randomly selected with equal probability of occurring anywhere inside the sample, as function of sample aspect ratio. Also shown in the figure is the strain ratio computed from the kMC model using actual simulation data with the annihilation direction passing through the grain CM. Note, the strain ratio is constant in time as the sample densifies. The shrinkage strain anisotropy resulting from the annihilation algorithm used in the kMC model is virtually identical to that obtained from the random direction model. Clearly, the current kMC annihilation model simulates performs annihilation in all direction with uniform frequency, which is leading to the shrinkage strain anisotropy in samples with aspect ratios not equal to unity. This means that while the present annihilation algorithm may simulate the microstructural evolution correctly, the macroscopic shrinkage of the sample is not simulated correctly.

\section{A new annihilation algorithm}
The annihilation algorithm must be modified to make the shrinkage strain isotropic for a sample of any aspect ratio. Ideally, a sample shrinking with isotropic strains during sintering would be characterized by the same local strain, $\alpha$.  Another way to describe this is that the strain at each position point in the sample relative to the sample center of mass would be the same at any time during sintering, $\alpha{}(t)$. For a continuous sample this would mean that the annihilation probability should be directly proportional to the distance from the current position point to the surface at any given point. The farther a point on the surface is from the current position point, the more annihilations will pass through it, and the more the distance will decrease. However, as we consider a voxel based model, moving a voxel on the surface at any given point will not result in a uniform reduction in distance, because the path will intersect a different number of voxels depending on where the surface point is located relative to the position point. In other words the length of the annihilation path divided by the number of voxels it passes through will be different for each annihilation path. This means that removing a voxel at the end of an annihilation path will not reduce the strain by one, but by a number between zero and $\sqrt{2}$, depending on the path.

An elegant solution to this problem of non-uniform strain on a digitized microstructure can be obtained by considering the requirements for isotropic strain in free sintering, which is
\begin{eqnarray}
\epsilon_x = \epsilon_y = \epsilon_z
\end{eqnarray}
The strain in the $x$ direction is given by
\begin{eqnarray}
\epsilon_x = \frac{L_x-L_{x0}}{L_{x0}},
\end{eqnarray}
where $L_x$ is the current length of the sample and $L_{x0}$ is the original length of the sample, respectively. The term $L_x-L_{x0}$ is the change in length due to shrinkage along the $x$-direction. As we are considering a digitized microstructure, the change in length can be quantified in the change in the number of voxels along the face in the $x$-direction. In the $x$-direction a sample face consists of $L_{y0} L_{z0}$ voxels for a rectangular sample with dimensions $L_{x0}\times{}L_{y0}\times{}L_{z0}$. This means that in order for the length $L_{x}$ to decrease by one, a total of $L_{y0} L_{z0}$ voxels must be annihilated. Thus the change in length is given by
\begin{eqnarray}
L_x-L_{x0} = \frac{n_x}{L_{y0} L_{z0}},
\end{eqnarray}
where $n_x$ is the number of annihilation event that terminate on the $x$-face, as each annihilation event shrinks that face by one voxel. This means that the strain in the $x$-direction is given by
\begin{eqnarray}
\epsilon_x = \frac{n_x}{L_{x0} L_{y0} L_{z0}}.
\end{eqnarray}
Following the same line of argument for the two other directions the requirement that a freely sintering sample has isotropic strain can be expressed as
\begin{eqnarray}
\frac{n_x}{L_{x0} L_{y0} L_{z0}} = \frac{n_y}{L_{x0} L_{y0} L_{z0}} = \frac{n_z}{L_{x0} L_{y0} L_{z0}}\; \Rightarrow \; n_x = n_y = n_z.
\end{eqnarray}
Thus in order for the strain to be isotropic the number of annihilation events terminating at each face of the sample must be the same, regardless of the sample dimensions.

Having determined this, the choice of a new algorithm is straight forward. Instead of picking a random direction along which the annihilation takes place, equal probability for annihilation terminating at the sample surfaces is assigned. A face is selected at random with each of the six faces having equal probability of being selected and a voxel on that face is then selected at random also with uniform probability. By selecting all faces with equal probability the number of annihilations events that pass through each face will be the same, and the strain will thus be isotropic in free sintering.
In other words, after the faces of the sample have been identified, the two-stage algorithm is:
\flushleft
\begin{enumerate}
\item Select a face on the sample
\item Choose a random point on that face and use that as the termination point for annihilation
\end{enumerate}
The faces of course have to be identified before the sintering starts. The number and relative orientation of the faces are assumed to remain constant through sintering. This means that any distortion or smoothing of e.g. corners is disregarded in selected the faces of the sample. This is a good approximation, as the overall shape change (disregarding uniform shrinkage) of a free sintering sample is usually small. For a spherical sample there is only one face, and the direction of annihilation would be equal in all direction, as it should.

\subsection{Strains for a microstructural sample}
To validate the new annihilation algorithm, we consider the strain and microstructural evolution for a powder sample consisting of packed spheres with a uniform distribution of radii between 5 and 10 voxels. This has previously been shown to be a reasonable particle size for sintering experiments \cite{Bjoerk_2012a}. The sample is formed by simulating the pouring of the spherical particles into a cubic container with dimensions $100\times{}200\times{}400$ voxels. The numerical code used to simulate this is the Large-scale Atomic/Molecular Massively Parallel Simulator (LAMMPS) code \cite{Plimpton_1995}, available as open source and developed by Sandia National Laboratories. Each powder particle is considered  a single crystal.

After the sample was formed, its sintering behavior was modeled using a simulation temperature of $k_B T= 1$ for grain growth and pore migration and $k_B T= 15$ for vacancy formation. The attempt frequencies were chosen in the ratio 1:1:5 for grain growth, pore migration and vacancy formation, respectively. These values, as previously shown \cite{Bjoerk_2012a}, result in realistic sintering behavior for a powder compact with grain boundary diffusion being the primarily mass transport mechanism, surface diffusion reshaping the pores to minimize the surface free energy and annihilation leading to densification. However, the chosen values are not meant to exactly replicate the shrinkage behavior observed for a given material system. Rather, we consider a general (random) representative case, which can be applied for any homogeneous material.

While the dimensions of the sample box was $100\times{}200\times{}400$ voxels, these will not be the true dimensions of the sample. The particles were poured from above into the sample box resulting in a situation where some of the particles at the top surface ``sticks out'' of the sample box after all particles have been poured. The particles that extend outside the sample box are removed from the sample. This leaves the top surface of the sample slightly more rough and smaller than the height of the sample box. Calculation of isotropic strains using the proposed algorithm requires the true sample dimensions to be found.

\begin{figure}[!t]
  \centering
\includegraphics[width=1\columnwidth]{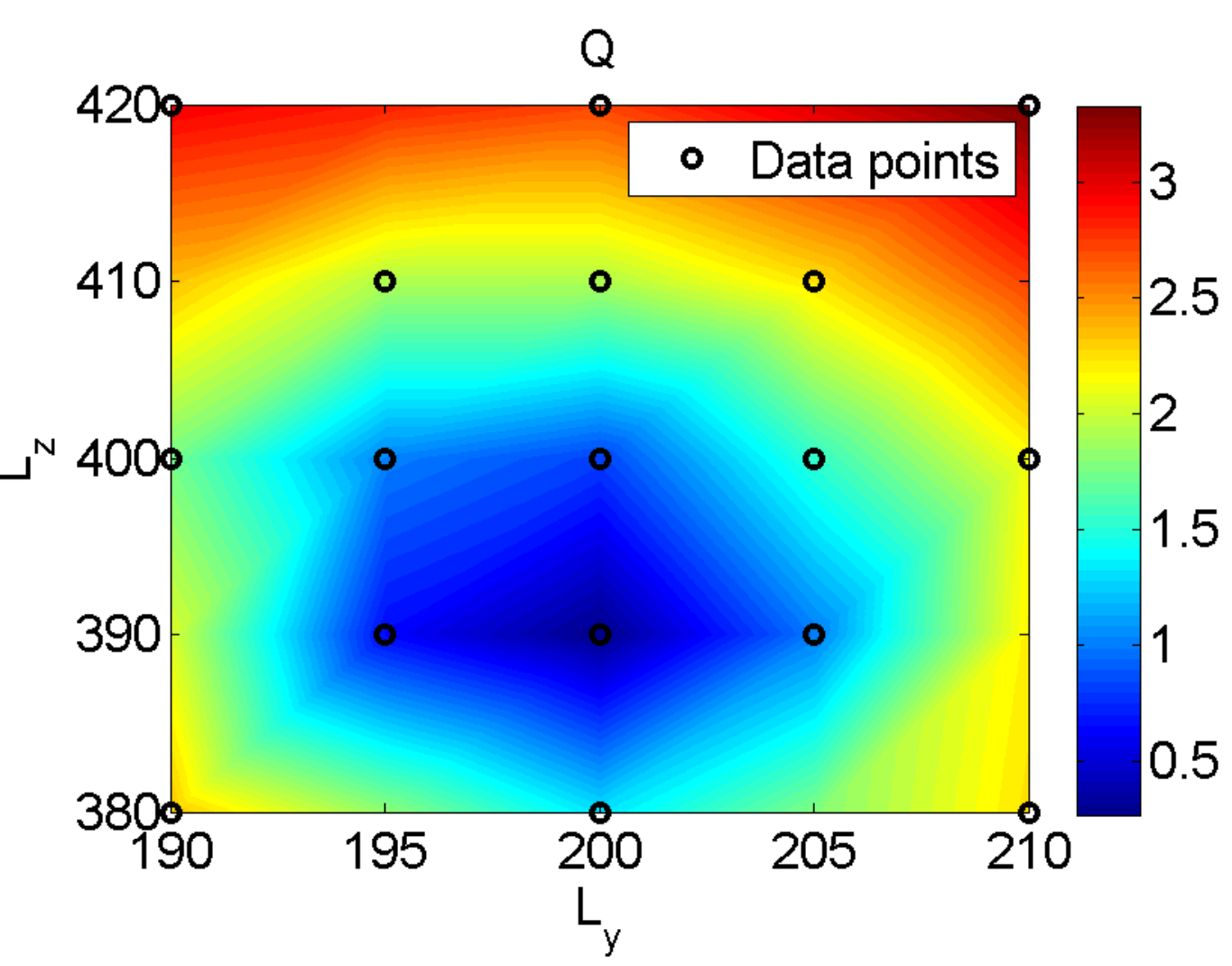}
\caption{The value of $Q=\sqrt{(\epsilon{}_x-\epsilon{}_y)^2+(\epsilon{}_x-\epsilon{}_z)^2}$ with $L_y$ and $L_z$. The lower the value of $Q$, the more isotropic the strain.}
    \label{Fig_Box_size_surface_plot}
\end{figure}

Shown in Fig. \ref{Fig_Box_size_surface_plot} is a surface plot of the value of $Q=\sqrt{(\epsilon{}_x-\epsilon{}_y)^2+(\epsilon{}_x-\epsilon{}_z)^2}$ after sintering, which is a direct measure of the how anisotropic the strain is for the different samples. For isotropic sintering $Q=0$. The value of $Q$ is shown as function of the assumed $y$ and $z$ dimensions of the samples, $L_y$ and $L_z$ respectively. As can be seen from the figure, isotropic strain is obtained by reducing the $L_z$ value slightly. This is expected, because, as explained previously, the particles sticking out out of the sample at the top surface were removed from the sample, reducing the height of the sample.

\begin{figure}[!t]
  \centering
\includegraphics[width=1\columnwidth]{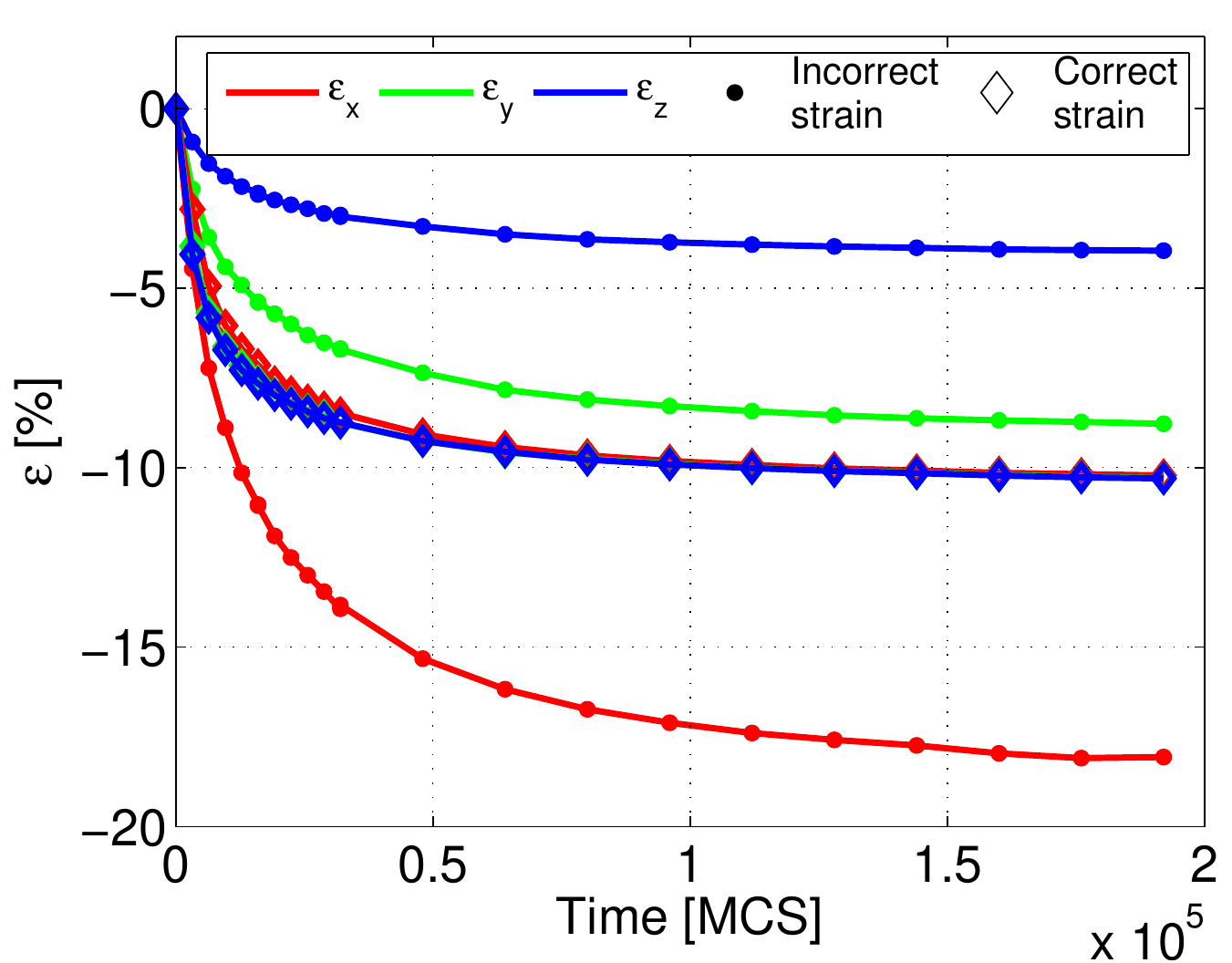}
\caption{The shrinkage strain, $\epsilon{}_x$, $\epsilon{}_y$ and $\epsilon{}_z$, as function of time in units of MCS for a simulation with the original and new annihilation algorithms.}
    \label{Fig_Compared_strain_fix_strains}
\end{figure}

Having found the true sample dimensions, the shrinkage strain can be determined as function of time, as the sample sinters. Time, in the model, is measured in Monte Carlo steps (MCSs), which are linearly related to real time \cite{Tikare_2010}. The linear coefficient depends on material parameters, and must be established through a comparison with experimental data for each material considered. However, as we consider a general case, this coefficient has not been determined. The strain in the three different directions are shown for both the old and the new algorithm in Fig. \ref{Fig_Compared_strain_fix_strains}. As can clearly be seen, the strain is isotropic with the new algorithm, while it is not so in the old algorithm with annihilation through the grain center of mass.

\begin{figure}[!t]
  \centering
\includegraphics[width=1\columnwidth]{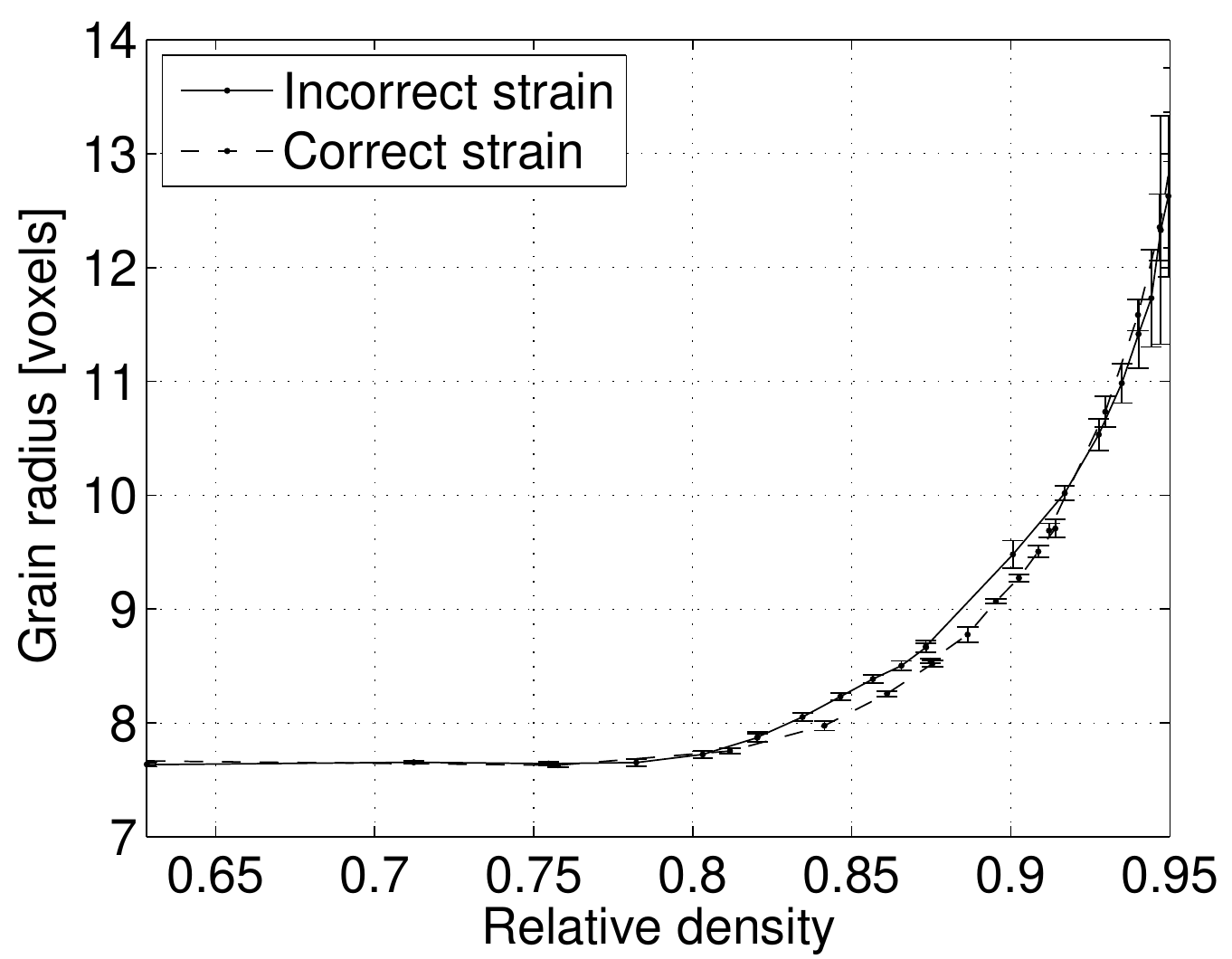}
\caption{The average grain size as function of the relative density for simulations with the original and modified annihilation algorithms. The error bars are the standard deviation of five simulations with different random seeds.}
    \label{Fig_Compared_strain_fix_microstructure}
\end{figure}

Comparing the microstructural evolution using the old and new annihilation algorithm is of critical importance. In Fig. \ref{Fig_Compared_strain_fix_microstructure} the average grain size as function of the relative density is shown for both the old and new annihilation algorithm. Five simulations with different random number seeds were run in order to estimate the uncertainty of the microstructural parameters. Although a small difference exist in microstructural evolution, the trends are clearly the same for both algorithms, and as such the new algorithm can be expected to compare equally well to experimental results as the old one did. This means that experimental validation of the kMC sintering model presented in Tikare et. al. (2010) \cite{Tikare_2010}, where the model was compared with experimental sintering data for Cu spheres obtained using microtomography, remains equally valid for the new annihilation algorithm presented above.

\section{Discussion}
The new algorithm presented above only works for samples with plane faces. If a simple problem such as the sintering of two or three spheres is considered, the above algorithm will not function as there are no plane faces of the studied system/sample. In this case the original algorithm will provide a more physically correct description of the sintering behavior.

If a sample is sintering under constraints, the strain will be very different from the isotropic case of free sintering. For such a sample the true deformation of the sample based on the constrains must be modeled. Handling annihilations events and paths and subsequent strains in this case will be considered in a future work.

\section{Conclusion}
The annihilation process and the resulting strains measured in the kinetic Monte Carlo (kMC) model of sintering have been analyzed. In the kMC model, densification occurs when an isolated pore site, termed a vacancy, is annihilated by collapsing a column of sites into the vacancy. The previously published algorithm for annihilation was shown to produce anisotropic strains for homogeneous samples with aspect ratios different from one. It is shown that the line direction biases shrinkage strains in proportion the compact dimension aspect ratios. In order for the strain to be isotropic it was shown that the number of annihilation paths terminating at each surface must be the same. A new algorithm based on this knowledge was proposed where a random face and a subsequent random point on that face is chosen with equal probabilities as the end point for an annihilation path. This was shown to result in isotropic strain for samples with any dimensions. The microstructural evolution was shown to be similar for the new and old algorithms, thus validating the new algorithm.

\section*{Acknowledgements}
The authors would like to thank the Danish Council for Independent Research Technology and Production Sciences (FTP) which is part of The Danish Agency for Science, Technology and Innovation (FI) (Project \# 09-072888) for sponsoring the OPTIMAC research work. Sandia National Laboratories is a multi-program laboratory managed and operated by Sandia Corporation, a wholly owned subsidiary of Lockheed Martin Corporation, for the U.S. Department of Energy's National Nuclear Security Administration under contract DE-AC04-94AL85000. The support by the National Science Foundation Division of Civil and Mechanical Systems and Manufacturing Innovations (NSF Grant CMMI 1234114) is also gratefully appreciated.

\end{document}